\documentstyle[12pt]{article}
\begin{document}
\begin{flushright}
Columbia preprint CU--TP--639
\end{flushright}
\vspace*{1cm}
\setcounter{footnote}{1}
\begin{center}
{\Large\bf Perturbative versus  Lattice QCD Energy Density Correlators
 at High Temperatures\footnote{ This work was supported by the Director,
 Office of Energy
Research, Division of Nuclear Physics of the Office of High Energy and Nuclear
Physics of the U.S. Department of Energy under Contract No.\
DE-FG-02-93ER-40764.}}
\\[1cm]
Dirk H.\ Rischke\footnote{Partially supported by the
Alexander von Humboldt--Stiftung under
the Feodor--Lynen program.} and Miklos Gyulassy \\ ~~ \\
{\small Physics Department, Pupin
Physics Laboratories, Columbia University} \\
{\small 538 W 120th Street, New
York, NY 10027, U.S.A.} \\ ~~ \\ ~~ \\
{\large August 1994}
\\[1cm]
\end{center}
\begin{abstract}
  Correlators of magnetic and electric field energy density are investigated
  for SU($N_c$) gauge theory at high temperatures $T$. At separations $z\leq
  2/T$ the correlators are shown to be dominated by a power--law behavior even
  for finite gluon screening masses.  This continuum behavior is well
  approximated on current $4\times 16^3$--lattices in the perturbative limit
  and leads to a considerable overestimate of screening masses deduced from
  fitting the lattice correlators with conventional exponential forms.  The use
  of extended sources and sinks to enhance the signal improves the situation
  for screening masses $m\gg T$ but leads to a largely uncontrolled error for
  masses less than $T$.  In fact, we show that recent lattice QCD data of
  Grossmann et al., from which a magnetic screening mass $m_M \approx 2.9 \, T$
  was deduced, may even be consistent with a vanishing actual magnetic
  screening mass.
\end{abstract}
\newpage
\section{Introduction and Conclusions}

In a recent lattice QCD study of SU(3) pure gauge theory \cite{grossman} the
plaquette--plaquette correlation functions were measured at high temperatures.
Below the deconfinement transition temperature ($T_c \sim 200$ MeV) the
observed exponential fall--off of those correlators was used to determine the
mass spectrum of glueballs. Above $T_c$ the two lowest masses deduced from the
decay of the correlators were interpreted as twice the electric and magnetic
screening mass in the gluon plasma.  The interesting result reported in
\cite{grossman} that we focus on is the claim that just above the critical
temperature, $T \approx 1.5 \, T_c$, the magnetic screening mass, $m_M/T
\approx 2.9 \pm 0.2$, is apparently twice as large as the electric screening
mass, $m_E/T \approx 1.4 \pm 0.2$.  Our interest in these results stems from
the sensitivity of certain transport properties and hence signatures of
quark--gluon plasmas produced in ultrarelativistic nuclear collisions
\cite{qm93,stoecker} to the ratio of these screening masses.  In Ref.\
\cite{selik} it was shown that the color relaxation time in a plasma is
\begin{equation}
  t_c = [3 \alpha_s T f(m_E/m_M )]^{-1}~,
\end{equation}
where $f(x)\approx \log(x)$ as $x\rightarrow \infty$.  This time
scale controls the lifetime of collective color excitations in the plasma such
as the plasmon.  Also next--to--leading--order corrections to the electric
screening mass depend self--consistently on the ratio $m_E/m_M$ \cite{rebhan}.
In lowest order perturbation theory, $m_E/T=g[(N_c + N_f/2)/3]^{1/2}$
 while $m_M/T=0$ \cite{weldon}.  The infrared magnetic singularities in higher
orders are however expected \cite{pisar} to produce a non--perturbative $m_M/T
\sim g^2$.  In the weak--coupling limit ($T\gg T_c$), $m_E/m_M \sim 1/g \gg
1$, and there is a natural ordering of the screening scales, $T\gg gT \gg
g^2T$.  However, at $T \approx 1.5\, T_c$, $m_E/m_M \approx 1/2$
\cite{grossman}, and the measured screening masses imply a reversal of that
perturbative ordering.  Since the temperatures accessible via nuclear
collisions are of the order of $2\, T_c$, this inverse ordering of screening
scales could have important phenomenological implications.  For example, it
implies a higher color conductivity which in turn could decrease the produced
transverse energy and the dilepton rate \cite{eskola}.

The screening masses are also of general interest as another test of the
viability of the perturbative picture of the quark--gluon plasma.  On the one
hand, the lattice entropy density \cite{christ}, color electric screening mass
\cite{karsch}, baryon susceptibility \cite{toussaint}, and spatial dependence
of higher mass meson and baryon correlators \cite{born} are all consistent with
the perturbative picture of the quark--gluon plasma, even near $T_c$.  On the
other hand, the temperature dependence of the energy density and pressure
deviate strongly from the ideal Stephan--Boltzmann behavior up to several times
$T_c$ \cite{engels,kogut,dhr}, the decay of the scalar and pseudoscalar meson
correlation functions cannot be reconciled with free propagation of the $q
\bar{q}$--pair \cite{born}, and $q\bar{q}$--wavefunctions behave as if the
quarks were strongly bound \cite{koch2}.

The recent measurement \cite{grossman} of $m_M \approx 2\, m_E$ poses a
further problem for the perturbative picture of the gluon plasma.  How can
static color magnetic fields be screened on such a small distance scale? In a
magnetic screening volume $4\pi/(3m_M^3)$ there is less than 1/2 gluon on the
average in the perturbative picture.  Non--perturbative mechanisms, such as a
very dense color magnetic monopole gas, would have to exist to account for that
large a magnetic screening mass.  However, there appears to be no consistent
effective QCD action that could generate such a monopole gas \cite{polonyi}.

In this work we aim to clarify how the lattice results of \cite{grossman} above
$T_c$ may translate into continuum physics.  To this end we assume that the
lattice spacing is small enough to allow the plaquette correlators to be
related to energy density correlators. The latter can be decomposed into
multi--gluon exchange diagrams. We furthermore assume that the dominant
contribution is given by the exchange of two dressed gluons. According to
standard arguments, this would give rise to a decay of the correlator governed
by twice the gluon screening mass, in agreement with the interpretation the
authors of \cite{grossman} have provided for their results. We first
investigate the two--gluon exchange diagram in the interaction--free case to
provide better insight on the screened case to be discussed subsequently.  For
distances $z \ll 1/T$, i.e., on momentum scales where temperature effects are
negligible, the correlators behave $\sim z^{-5}$, as expected from dimensional
arguments.  For $z\geq 1/2T$ this behavior is changed to a decay $\sim z^{-4}$
due to finite temperature effects.  Then we study the effect of self--energy
insertions.  We show that the structure of the electric correlator is
considerably more complex than that of the magnetic since the former mixes the
longitudinal and transverse polarization effects.  In both cases, for $z\leq
2/T$ the correlators are found to be dominated by the power--law behavior
encountered in the interaction--free case and not by the exponential decay
expected for gluons with a finite screening mass.

We then calculate the correlators on a lattice in the weak--coupling limit and
show that the continuum limit is well approximated on a $4\times
16^3$--lattice. Consequently, the power--law behavior is still prominent
\cite{berg} and leads to an overestimate of the masses deduced from
conventional exponential--type fit functions. The power--law behavior is caused
by the point source and sink of the correlator. However, in \cite{grossman}
fuzzed links \cite{fuzz} were used which have the effect of ``smearing'' out
the sources and sinks and partially remove the power--law enhancement at small
distances. We introduce a model for fuzzed links which incorporates the
smearing and show that the relative residual error in determining the screening
masses indeed becomes small if the latter are large compared to $T$.  However,
for a {\em vanishing\/} screening mass this fuzzing model still leads to an
apparent {\em finite\/} mass.  In fact, we show that the current lattice data
\cite{grossman} may even be consistent with a vanishing magnetic screening
mass~!  The screening masses obtained from energy density correlators are
therefore inconclusive and require confirmation from an independent measurement
of e.g.\ the gluon propagator \cite{private2}.

The remaining sections are organized as follows. In Section 2 we review in more
detail the results of Ref.\ \cite{grossman} and relate them to our
investigations.  In Section 3 the energy density correlators are decomposed in
terms of multi--gluon exchange. Section 4 comprises a discussion of the
interaction--free case. In Section 5 we discuss the correlators for dressed
gluons. Section 6 is devoted to the transcription of the perturbative continuum
results onto the lattice.  In Section 7 we investigate a model for fuzzed links
and their effect on the extraction of screening masses.

We use natural units $\hbar=c=1$ and, except for Section 6 and 7, the
Minkowski metric $g^{\mu \nu} = {\rm diag}(+,-,-,-)$. This implies $x^{\mu}
= (-i \tau, {\bf x})$ and $k^{\mu} = (-i\omega_n,{\bf k})$, with the gluonic
Matsubara frequencies $\omega_n = 2 \pi n T$.

\section{Plaquette correlators in lattice QCD}

In the perturbative picture of a quark--gluon plasma as a weakly interacting
system, $q\bar{q}$-- (or meson) correlators are essentially given by a single
$q\bar{q}$--loop, since the mutual interaction between the quark and the
antiquark is screened by the hot environment. Also, a quark acquires a
dynamical mass $m_q \sim gT$ \cite{quarkmass}.  The correlator is roughly
proportional to the square of the (dressed) quark propagator \cite{shuryak}.
For (large) spatial distances, the decay of the latter is governed by the
lowest fermionic ``energy'' $[(\pi T)^2 + m_q^2]^{1/2}$ ($\pi T$ being the
smallest fermionic Matsubara frequency).  Since $\pi^2 \gg g^2$ we may
neglect the quark mass as compared to the Matsubara frequency \cite{footn6} and
the spatial meson correlator decays like $\exp \{ - 2 \pi zT\}$ \cite{friman}.
After correcting for lattice artefacts, the lattice data for vector and axial
vector meson are found to be in good agreement with this picture \cite{born}.
The scalar/pseudoscalar channel is, however, an important exception in that it
only decays $\sim \exp \{ - \pi zT \}$. This points towards a strong residual
interaction in this channel, see e.g.\ \cite{goksch,schaefer}.

However, as the study of $q \bar{q}$--wavefunctions has revealed \cite{koch2},
{\em even\/} in the vector channel the quark--antiquark pair may be strongly
bound.  The reason why the corresponding binding energy does not show up in the
mass extracted from the correlator is the same why we could neglect the quark
mass in our above consideration: it is negligible as compared to $2 \pi T$
which originates from the lowest fermionic Matsubara frequency \cite{koch2}. It
is therefore difficult to extract either the residual interaction effects or
dynamical quark masses from spatial $q\bar{q}$--correlation functions
\cite{boyd}.

In view of this, the study of glueballs is of special interest: since the
lowest Matsubara frequency for gluons is zero, the measurement of glueball
correlators is not impaired by the above mentioned effect.  They may thus
provide more detailed information on residual interactions and the gluon
screening mass.  In \cite{grossman} the irreducible representations for the
glueball states on the lattice were identified and their masses measured via
the decay of the corresponding spatial correlation functions \cite{berg}. It
was found that above $T_c$ the lowest mass state no longer corresponds to a
glueball with definite quantum numbers. Rather, it belongs to the (connected)
correlator between plaquette operators of the form
\begin{equation} \label{Pt}
  P_t \equiv P_{0x} + P_{0y}~,
\end{equation}
where
$$
P_{0i} =\frac{1}{2N_c}~{\rm Tr} \left[ U_0 U_i U^{\dagger}_0
U^{\dagger}_i \right]
$$
is the elementary temporal plaquette with two spatial links $U_i$
and two temporal links $U_0$,
$$
U_{\mu} = \exp \left\{ iga {\cal A}_{\mu} \right\}~,
$$
with ${\cal A}_{\mu} \equiv A^b_{\mu} G^b$, and the generators of
SU($N_c$) $G^b,\, b = 1,...,N_c^2-1$.  The lattice spacing $a$ is a dynamical
quantity in the sense that it decreases with decreasing coupling $g$ according
to the QCD $\beta$--function \cite{bbeta}. Since $g$ decreases with increasing
temperature, at sufficiently high $T$ the lattice spacing should be small
enough to permit an expansion of the plaquette operator in powers of $a$. To
lowest (non--trivial) order \cite{footn2} in $a$, the plaquette operator
(\ref{Pt}) is therefore proportional to
\begin{equation} \label{plaqt}
  P_t \sim {\rm Tr} \left[ {\cal E}_x^2 + {\cal E}_y^2 \right]~,
\end{equation} where ${\cal E}_i \equiv {\cal F}_{i0} = F^b_{i0} G^b = E^b_i
G^b$ is the $i$th component of the electric field strength (${\cal F}_{\mu
  \nu}$ is the field--strength tensor). Thus, (\ref{Pt}) is proportional to the
color electric field energy density.  As will be shown in detail in Section 3,
the field energy density correlator can be decomposed in terms of multi--gluon
exchange diagrams, the simplest of which is a one--loop, two--gluon exchange
diagram.  In the perturbative picture of the gluon plasma, this diagram is
expected to give the dominant contribution to the energy density correlator,
since mutual interactions between gluons are screened by the hot environment.
In Section 5 we will show that in this case the mass determining the
exponential decay of the energy density correlator is twice the gluon screening
mass.  It was found in \cite{grossman} that the mass extracted from the
correlator between the $P_t$ is consistent with the mass extracted from the
Polyakov--loop correlation function, which for $T \approx 1.5 \, T_c$ agrees
well with the perturbative result $2\, m_E$. This suggests that indeed the
perturbative picture of the gluon plasma is valid, i.e., that the one--loop,
two--gluon exchange diagram constitutes the dominant contribution to the
$P_t$--correlator at $T \approx 1.5\, T_c$.  In our following continuum
analysis, this will be our main working {\em assumption}.  However, we
emphasize that there is no rigorous proof that the respective temperature
$T\approx 1.5\, T_c$ is sufficiently high and, consequently, the lattice
spacing $a$ sufficiently small so that the energy density correlators
accurately reflect the plaquette correlators, and that the dominant
contribution to the energy density correlators is the one--loop, two--gluon
exchange.

The second lowest mass extracted from the spectrum of all glueball--like
excitations above $T_c$ was identified to belong essentially to the connected
correlator between elementary spatial plaquettes
\begin{equation}
  P_s \equiv P_{xy}~,
\end{equation}
which, to lowest non--trivial order in $a$, are
proportional to the magnetic field energy density in $z$--direction:
\begin{equation}
  P_s \sim {\rm Tr}~{\cal B}^2_z~,
\end{equation}
${\cal B}_z \equiv {\cal F}_{xy} = F^b_{xy} G^b = B^b_z G^b$.
In analogy to the electric correlator, the authors of \cite{grossman}
interpreted this screening mass as twice the non--perturbative magnetic
screening mass $m_M$.

\section{Energy density correlators in terms of multi--gluon
exchange}

As indicated in the last section, if one assumes that the dominant contribution
to the energy density correlators is given by two--gluon exchange, then this
allows an interpretation of the mass determined from the decay of the
correlators as twice the gluon screening mass. In order to clarify this, we
decompose the energy density correlators into multi--gluon exchange diagrams.
We start with the magnetic correlator which has the simpler structure.  The
magnetic point--to--point correlator between $r \equiv (-i \tau,{\bf x})$ and
$r' \equiv (-i \tau', {\bf x}')$ reads
\begin{eqnarray}
  M(r,r') & \equiv & \left\langle {\rm Tr}~{\cal B}^2_z(r)~{\rm
    Tr}~{\cal B}^2_z(r') \right\rangle_c \nonumber \\ & \equiv & {\rm
    Tr} [ G^a G^b ]~{\rm Tr} [ G^c G^d ]~\left\langle F_{xy}^a (r)
  F_{xy}^b (r) F_{xy}^c (r') F_{xy}^d (r') \right\rangle_c~.
\end{eqnarray}
The subscript ``c'' at the ensemble average brackets indicates
that only the connected part of the respective expression is to be taken into
account. Each color trace yields a factor 1/2 and a Kronecker delta in the
color indices. Inserting the definition of the non--Abelian field strength
$$
F_{\mu \nu}^a \equiv \partial_{\mu}
A_{\nu}^a - \partial_{\nu} A_{\mu}^a - g f^{abc} A_{\mu}^b A_{\nu}^c~,
$$
we arrive at the following  decomposition
\begin{equation}
  M(r,r') \equiv M^{(0)}(r,r') + g^2~M^{(1)}(r,r') + g^4~M^{(2)}(r,r')
\end{equation}
with
\begin{eqnarray} \label{m0r}
  M^{(0)}(r,r') & = & \frac{1}{4} \left\langle \left[ \partial_x
  A_y^a(r) -
\partial_y A_x^a(r) \right]^2 \left[ \partial_{x'} A_y^b(r') -
\partial_{y'} A_x^b(r') \right]^2 \right\rangle_c~, \\
M^{(1)}(r,r') & = & \frac{1}{4} f^{acd} f^{aef} \left\{ \left\langle
\left[ \partial_x A_y^b(r)-\partial_y A_x^b(r) \right]^2 A_x^c(r')
A_y^d(r') A_x^e(r') A_y^f(r') \right\rangle_c \right. \nonumber \\ & &
\left.  \hspace{1.7cm} + \left\langle \left[ \partial_{x'}
A_y^b(r')-\partial_{y'} A_x^b(r') \right]^2 A_x^c(r) A_y^d(r) A_x^e(r)
A_y^f(r) \right\rangle_c \right\} \label{m1r} \\ +~f^{acd} f^{bef} & &
\!\!\!\!\!\!\!\!\!\!\! \left\langle \left[ \partial_x
A_y^a(r)-\partial_y A_x^a(r) \right] A_x^c(r) A_y^d(r) \left[
\partial_{x'} A_y^b(r')-\partial_{y'} A_x^b(r') \right] A_x^e(r')
A_y^f(r') \right\rangle_c~, \nonumber \\
M^{(2)}(r,r') & = &
\frac{1}{4}~f^{acd} f^{agh} f^{bef} f^{bij} \left\langle A_x^c(r)
A_y^d(r) A_x^g(r) A_y^h(r) A_x^e(r') A_y^f(r') A_x^i(r') A_y^j(r')
\right\rangle_c\label{m2r}
\end{eqnarray}
In order to calculate the ensemble averages we Fourier--transform the
gluon fields according to
\begin{equation}
  A_{\mu}^a(r) = \frac{1}{\sqrt{TV}} \sum_k e^{-i k \cdot r}
  A_{\mu}^a(k)~.
\end{equation}
The normalization makes the Fourier amplitudes $A_{\mu}^a(k)$
dimensionless numbers. In the thermodynamic limit we have as usual
$$
\sum_k \equiv \sum_{k^0} \sum_{\bf k} \rightarrow \sum_{k^0}~V \int
\frac{{\rm d}^3 {\bf k}}{(2 \pi)^3}~~~~(V \rightarrow \infty)~.
$$
An explicit calculation of the $n$--gluon amplitudes in
(\ref{m0r}--\ref{m2r}) is possible if we assume that the average is taken over
a Gaussian ensemble, i.e., with an action of the form
\begin{equation} \label{action}
  S[A] \equiv - \frac{1}{2 T^2} \sum_{k,k'} A_{\mu}^a(k)~( {\cal D}^{-1}
  )^{\mu \nu}_{ab} (k,k')~A_{\nu}^b(k')~.
\end{equation}
Note that ${\cal D}$ contains a gauge fixing term.  The factor
$1/T^2$ is conventional in order that ${\cal D}^{\mu \nu}_{ab}$ agrees with the
standard definition of the gluon propagator.  The assumption of a Gaussian
ensemble leads to the factorisation of an arbitrary $n$--gluon amplitude into a
sum over products of all possible pairings of two--gluon amplitudes, i.e.,
gluon propagators, \cite{glimm} via
\begin{eqnarray} \label{pair}
  \langle A_1 ... A_n \rangle & = & \sum_{pairs}~\langle A_{i_1} A_{i_2}
  \rangle~...~\langle A_{i_{n-1}} A_{i_n} \rangle \\ \langle
  A_{\mu}^a(k) A_{\nu}^b(k') \rangle & = & T^2~{\cal D}_{\mu \nu}^{ab}
  (k,k')~.
\end{eqnarray}
In the first line we used a collective index for
momentum, Minkowski--, and color--indices.  We furthermore assume
translation invariance and global color neutrality so that
$$
{\cal D}^{\mu \nu}_{ab}
(k,k') \equiv {\cal D}^{\mu \nu} (k)~\delta_{ab}~\delta^{(4)}_{k,-k'}~.
$$
With the abbreviation $T/V \sum_k \equiv \int_k$ we obtain
\begin{eqnarray}
  M^{(0)} (r,0) & = & \frac{N_c^2 - 1}{2} \int_{k,l} e^{-i(k+l) \cdot r}
  \left[ k_x^2~{\cal D}^{yy}(k) - 2~k_x k_y~{\cal D}^{yx}(k) +
  k_y^2~{\cal D}^{xx}(k) \right] \nonumber \\
& & \hspace{3.5cm} \times
  \left[ l_x^2~{\cal D}^{yy}(l) - 2~l_x l_y~{\cal D}^{yx}(l) +
  l_y^2~{\cal D}^{xx}(l) \right]~,\label{m0}
\end{eqnarray}
and similar expressions for $M^{(1)}$ and $M^{(2)}$.  Each term in eq.\
(\ref{m0}) has the graphical interpretation shown in Figs.\ 1a and 1d.
We represent the origin O and the point $r$ by ``vertices'' (without the usual
factor $g$~!) and gluon propagators between them by full lines (for free
gluons, i.e., ${\cal D} \equiv {\cal D}_0$ in (\ref{m0}), we use simple full
lines as in Fig.\ 1a, for dressed gluons, we insert a self--energy ``bubble''
as in Fig.\ 1d).  A cross on the line corresponds to one extra momentum factor
or partial derivative.  There are three topologically distinct two--loop graphs
contributing to $M^{(1)}$ and two three--loop graphs contributing to $M^{(2)}$.
Examples of these are illustrated in Figs.\ 1b and 1c for ${\cal D} \equiv
{\cal D}_0$.  However, we do not consider those further, in accordance with our
working assumption that the one--loop, two--gluon exchange represents the
dominant contribution to the energy density correlators.

Other diagrams which will not be considered in the following arise from
non--Gaussian fluctuations, which we have neglected in the factorisation
(\ref{pair}), see for instance Fig.\ 1e. This generic ladder diagram
could lead to bound states. In the perturbative picture
of the gluon plasma, such a diagram would be suppressed
since the ladder interactions are screened by the hot environment.
Nevertheless, in view of the analysis of hadronic wavefunctions
\cite{koch2} one cannot a priori rule out that these graphs are not important
near $T_c$. Then, however, the exponential decay of the correlators would be
determined by the mass of the exchanged bound gluon pair rather than by twice
the screening mass of the two dressed gluons of Fig.\ 1d.  This points to an
intrinsic ambiguity in the interpretation of the energy density correlators.
Fig.\ 1f will be discussed in Section 7.

The electric field energy density correlator,
\begin{equation}
  E(r,r') \equiv \left\langle {\rm Tr} \left[ {\cal E}^2_x(r) + {\cal
    E}_y^2(r)\right]~{\rm Tr} \left[ {\cal E}^2_x(r') + {\cal
    E}_y^2(r')\right] \right\rangle_c~,
\end{equation}
can be treated in a completely analogous way. As in the magnetic case, our
subsequent discussion will be restricted to the one-loop term
\begin{eqnarray}
E^{(0)} (r,0) & \!\!\!= & \!\!\!\! \frac{N_c^2 - 1}{2} \int_{k,l} \!
e^{-i(k+l) \cdot r} \!\!\! \sum_{i,j=x,y} \!\! \left[ (k^0)^2 {\cal
D}^{ij}(k) - k^0 k^j {\cal D}^{i0}(k) - k^i k^0 {\cal D}^{0j}(k) + k^i
k^j {\cal D}^{00}(k) \right] \nonumber \\
&   & \hspace{3.55cm} \times
\left[ (l^0)^2 {\cal D}^{ij}(l) - l^0 l^j {\cal D}^{i0}(l) - l^i l^0
{\cal D}^{0j}(l) + l^i l^j {\cal D}^{00}(l) \right]. \label{e0}
\end{eqnarray}
We finally project the correlators onto a state with
$\omega_n=k_x=k_y=0$.  This is achieved by integrating them over the transverse
$\tau-x-y$--plane at $r$, yielding the {\em spatial\/} magnetic correlator
\begin{equation} \label{project}
  M(z) \equiv \int_0^{1/T} {\rm d}\tau \int_{-L/2}^{L/2} {\rm d}x
  \int_{-L/2}^{L/2} {\rm d}y~M(r,0)~,
\end{equation}
and similarly for the spatial electric correlator. The
integration over the transverse plane eliminates three of the momentum sums in
(\ref{m0}) and (\ref{e0}) via the identity
\begin{equation} \label{complete}
  \int_0^{1/T} {\rm d}\tau \int_{-L/2}^{L/2} {\rm d}x \int_{-L/2}^{L/2}
  {\rm d}y~e^{-ik \cdot r} = e^{i k^z z}
  \frac{L^2}{T}~\delta_{k^0,0}~\delta_{k^x,0}~\delta_{k^y,0}~.
\end{equation}
Explicit expressions will be provided in the next sections when
the form of the gluon pro\-pa\-gator is specified.

\section{The interaction--free case}

In this section we discuss the spatial correlators for the case of vanishing
coupling $g=0$ (indicated by subscript 0 at the respective quantities). In this
case, the one--loop graph of Fig.\ 1a is the only term contributing to the
correlators and is manifestly gauge invariant. Furthermore, ${\cal D}^{\mu\nu}$
in eqs.\ (\ref{m0}, \ref{e0}) reduces to the free gluon propagator which in the
covariant gauge is given by
\begin{equation} \label{FP}
  {\cal D}_0^{\mu \nu}(k) = \frac{g^{\mu \nu}}{k^2} - \alpha~\frac{k^{\mu}
  k^{\nu}}{(k^2)^2}~.
\end{equation}
For the integrand in (\ref{m0}) we need the expression
\begin{equation} \label{free}
  k_x^2~{\cal D}^{yy}_0(k) - 2~k_x k_y~{\cal D}^{yx}_0(k) + k_y^2~{\cal
    D}^{xx}_0(k) = \frac{k_x^2 + k_y^2}{-k^2} \equiv
    \frac{k_{\perp}^2}{\omega_n^2 +k_{\perp}^2 +k_z^2}~,
\end{equation}
where $k_{\perp}^2 \equiv k_x^2 + k_y^2$. Note that the gauge
parameter drops out, as expected. We now insert (\ref{m0}) into
(\ref{project}) and employ (\ref{complete}). Furthermore, in the
thermodynamic limit
$$
\frac{1}{L} \sum_{k^i} \equiv \int_{-\infty}^{\infty} \frac{{\rm
    d}k^i}{2 \pi}~,
$$
and \cite{gradst1}
\begin{equation} \label{FT}
  \int_{-\infty}^{\infty} \frac{{\rm d}k^z}{2 \pi}~e^{ik^z
    z}~\frac{1}{\tilde{E}^2 + k_z^2} \equiv
  \frac{1}{2\tilde{E}}~e^{-\tilde{E} z}~,~~~~z>0~,
\end{equation}
where we have introduced the abbreviation $\tilde{E}^2
\equiv \omega_n^2 + k_{\perp}^2$. We thus arrive at
\begin{equation} \label{23}
  M^{(0)}_0(z) = \frac{N_c^2-1}{2}~T \sum_n \int_0^{\infty}
  \frac{ {\rm d}k_{\perp} k_{\perp}}{2 \pi}~\frac{k_{\perp}^4}{4
    \tilde{E}^2}~e^{-2 \tilde{E}z}~.
\end{equation}
We remark that the exponential decay factor in (\ref{FT})
and in the integrand of (\ref{23}) is generic for spatial correlators.
The smallest exponential factor (and thus the dominant term at large
distances $z$) is that for the lowest Matsubara mode and zero transverse
momentum.

For further purposes it is convenient to first consider arbitrary (even) powers
of transverse momentum in the nominator of the integrand in (\ref{23}) and to
write $\tilde{E}^2 \equiv \mu^2 + k_{\perp}^2$.  In our case, $\mu^2
=\omega_n^2$, but in general, when there is also a mass term in the
propagator, $\mu^2= \omega_n^2 + m^2$.  Then, the integral reads
\begin{equation}
  \frac{1}{8 \pi}\int_0^{\infty} {\rm d}k_{\perp}
  k_{\perp}~\frac{k_{\perp}^{2j}}{\mu^2 + k_{\perp}^2}~e^{-2z
    \sqrt{\mu^2 + k_{\perp}^2}} \equiv
  \frac{1}{8\pi}~\frac{1}{(2z)^{2j}}~{\cal C}_j(2z |\mu|)~,
\end{equation}
where the functions
\begin{equation}
  {\cal C}_j(u) \equiv \int_{u}^{\infty} {\rm
    d}y~\frac{(y^2-u^2)^j}{y}~e^{-y}
\end{equation}
are related to the exponential integral. We will only need \cite{gradst2}:
\begin{eqnarray}
{\cal C}_0(u) & = & - {\rm Ei}(-u) \equiv e^{-u} \left( \frac{1}{u}
- {\cal J}(u) \right)~,\\
{\cal C}_1(u) & = & e^{-u} (1+u) - u^2~{\cal C}_0(u) \equiv e^{-u}
\left( 1 + u^2 {\cal J}(u) \right)~, \\
{\cal C}_2(u) & = & e^{-u}(6 + 6u + 3u^2 + u^3) - 2u^2~e^{-u}(1+u)
+ u^4~{\cal C}_0(u) \nonumber \\
& = & e^{-u} \left( 6 +6u + u^2 - u^4~{\cal J}(u) \right)~, \label{c2}
\end{eqnarray}
where
\begin{equation}
  {\cal J}(u) \equiv \int_0^{\infty} {\rm d}t~e^{-t}~\frac{1}{(u+t)^2}~.
\end{equation}
Note that
\begin{eqnarray} \label{uinf}
{\cal C}_j (u) & \sim & e^{-u}~,~~{\rm for}~u > 0~,\\
{\cal C}_1(0) & = & 1~,~{\cal C}_2(0)=6~,~~{\rm and}~~{\cal C}_0(u)
\sim \ln u~~~{\rm for}~u \rightarrow 0~. \label{u0}
\end{eqnarray}
In eq.\ (\ref{23}), $\mu \equiv \omega_n$, and thus
\begin{eqnarray}
\frac{M^{(0)}_0}{T^5}(zT) & = & \frac{N_c^2-1}{16\pi}~
\frac{1}{(2zT)^4}~\sum_{n=-\infty}^{\infty}~{\cal C}_2(2z
|\omega_n|) \nonumber \\
& = & \frac{N_c^2-1}{16\pi}~\frac{1}{(2zT)^4}~\left\{~{\cal C}_2(0)
+ 2 \sum_{n=1}^{\infty}~{\cal C}_2(4\pi n zT) \right\}~.\label{m00}
\end{eqnarray}
With eq.\ (\ref{c2}), the sum over $n$ can be further simplified
using identities for the geometrical progression and its derivatives. For the
following discussion, however, the form (\ref{m00}) is more convenient.

Let us first estimate the behavior of the correlator for different regions of
$zT$. For very small distances $zT \ll 1$, one expects that temperature effects
do not play a role and that the behavior of the correlator is solely determined
by dimensional arguments, as it would also be the case at $T=0$.  Since $M(r,0)
\sim r^{-8}$, we conclude that due to the integration over the
$\tau-x-y$--plane in (\ref{project}) $M(z)$ must be $\sim z^{-5}$ \cite{berg}.
Indeed, taking that term in (\ref{c2}), which is dominant for small $zT$, and
employing the geometrical progression one easily shows that the sum over
Matsubara frequencies $n\neq0$ in (\ref{m00}) diverges as $1/zT$, and hence
$M_0^{(0)}/T^5$ has the expected $(zT)^{-5}$--singularity.

At larger $zT$ the contributions of the non--zero Matsubara modes decay
exponentially according to (\ref{uinf}). Therefore, by virtue of (\ref{u0}),
the first term in curly brackets (corresponding to $n=0$) is the dominant term
for large $zT$. This changes the small distance $(zT)^{-5}$--behavior to a
decay proportional to $(zT)^{-4}$. Thus, finite temperature enhances long
distance correlations by one power of the distance as compared to the
$T=0$--case (in our case, from $z^{-5}$ to $z^{-4}$). This is a generic feature
for {\em all\/} correlation functions at finite temperature and is essentially
caused by the replacement of the time integration by $T$ times a Matsubara sum
in finite--temperature field theory \cite{kapusta}.

These expectations are confirmed in Fig.\ 2a where we have plotted the full
result (\ref{m00}) and the contribution of the zero mode as a function of $zT$.
For small $zT$, the non--zero Matsubara modes ($\sim (zT)^{-5}$ for $zT
\rightarrow 0$) dominate over the $n=0$--term (dotted line) which is only $\sim
(zT)^{-4}$, but already for $zT\approx 1/2$, they become negligibly small due
to the exponential suppression, and the decay of the correlator is $\sim
(zT)^{-4}$.

Note that since ${\cal C}_2(0)$ is {\em constant}, cf.\ eq.\ (\ref{u0}), the
decay of the spatial correlator in the interaction--free case is {\em
  polynomial}, $\sim (zT)^{-4}$ for $zT \geq 1/2$. However, as will be
discussed in the next section, in the case of screening interactions the gluon
will acquire a mass $m$. As mentioned above, this has the effect that $\mu =
\omega_n \rightarrow (\omega_n^2 + m^2)^{1/2}$, and the $n=0$--term in
(\ref{m00}) becomes ${\cal C}_2(2 z m)$.  Consequently, it {\em also decays
  exponentially\/} on account of (\ref{uinf}).  This exponential decay is
superimposed on the polynomial $(zT)^{-4}$--decay and of course dominates for
large $zT$.  At distances where $zT$ as well as $zm$ are small, however, the
scales set by the temperature or the mass, respectively, cannot play a role and
the correlator must again exhibit a $z^{-5}$--singularity as expected from
dimensional arguments.

Employing eq.\ (\ref{e0}), the analogue of (\ref{project}) for the
electric case, and eqs.\ (\ref{FP}, \ref{FT}), the electric correlator
reads
\begin{equation}
  E^{(0)}_0(z) = \frac{N_c^2-1}{2}~T \sum_{n} \int_0^{\infty}
  \frac{{\rm d}k_{\perp} k_{\perp}}{2 \pi}~\frac{ k_{\perp}^4 + 2\, \omega_n^2
  \tilde{E}^2 }{4\tilde{E}^2}~e^{-2 \tilde{E} z}~,
\end{equation}
and can therefore be partly expressed in terms of the
magnetic correlator,
\begin{eqnarray} \label{e00}
\frac{E^{(0)}_0}{T^5} (zT) & \equiv & \frac{M^{(0)}_0}{T^5}(zT) \\
& + & \frac{N_c^2-1}{4\pi}~\frac{1}{(2zT)^4}~\sum_{n=1}^{\infty}~(4
\pi n zT)^2 \left\{~{\cal C}_1(4 \pi nzT) +(4 \pi n zT)^2~{\cal C}_0(4
\pi n zT) \right\}~. \nonumber
\end{eqnarray}
Again, the sum can be analytically performed using identities
for the geometrical progression.

In Fig.\ 2b we show the electric correlator in comparison to the magnetic
correlator. As expected from (\ref{uinf}), the additional terms in the electric
correlator (\ref{e00}) are negligible for $zT \geq1$ and only influence the
domain of $0.1 \leq zT \leq 1$. For very small $zT \leq 0.1$, however, the
electric correlator becomes equal to the magnetic (and thus has the same
dominant singularity $\sim (zT)^{-5}$).  The reason is that electric and
magnetic fields are indistinguishable in Euclidean field theory at $T=0$ or at
distances where the scale set by the temperature becomes irrelevant,
respectively.

\section{Correlators in the case of dressed propagators}

In this section we study the effect of the hot environment on the
one--loop energy density correlators. To this end we assume that the two gluon
propagators connecting O and $r$ acquire a self--energy modification,
see Fig.\ 1d. The ``dressed'' propagator reads
(in covariant gauge) \cite{weldon}
\begin{equation} \label{D}
  {\cal D}^{\mu \nu}(k) = \frac{{\cal P}^{\mu \nu}}{k^2 - \pi_T} +
  \frac{{\cal Q}^{\mu \nu}}{k^2- \pi_L} + (1- \alpha)~\frac{k^{\mu}
    k^{\nu}}{(k^2)^2}~,
\end{equation}
with the two--space projectors
\begin{eqnarray}
{\cal P}^{\mu \nu} & \equiv & \Delta^{\mu \nu} + \frac{\tilde{k}^{\mu}
\tilde{k}^{\nu}}{{\bf k}^2}~, \\ {\cal Q}^{\mu \nu} & \equiv & g^{\mu
\nu} - \frac{k^{\mu} k^{\nu}}{k^2} - {\cal P}^{\mu \nu}~,
\end{eqnarray}
where $\Delta^{\mu \nu} \equiv g^{\mu \nu} - u^{\mu}
u^{\nu},\, \tilde{k}^{\mu} \equiv \Delta^{\mu \nu} k_{\nu}$.  $u^{\mu}$ is
the (collective) 4--velocity of the matter under consideration.

We now consider the combination of $k^i$ and ${\cal D}^{ij}$ appearing in
eq.\ (\ref{m0}) in the rest frame of matter, $u^{\mu} = (1,0,0,0)$. We
find the simple gauge invariant result
\begin{equation} \label{35}
  k_x^2~{\cal D}^{yy}(k) - 2~k_x k_y~{\cal D}^{yx}(k) + k_y^2~ {\cal D}^{xx}(k)
  = \frac{k_x^2 + k_y^2}{-k^2 + \pi_T} = \frac{k_{\perp}^2}{\omega_n^2 +
    k_{\perp}^2 + k_z^2+\pi_T}~,
\end{equation}
which should be compared with (\ref{free}). The only
modification thus far is the appearance of the ``gluon mass'' $\sqrt{\pi_T}$.
In the general case, $\pi_T$ depends non--trivially on $k^{\mu}$.  This may be
seen considering for instance the gauge--invariant part of the one--loop gluon
self--energy \cite{weldon,BP}. Only for $k^{\mu} = (0,{\bf k})$, i.e., for
the lowest Matsubara frequency, the dependence is trivial: $\pi_T(0,{\bf k})
\equiv 0$ for all ${\bf k}$, indicating the absence of screening of static
magnetic fields in first order perturbation theory.  Static magnetic fields are
screened in higher order in $g$, but this is not perturbatively calculable. On
the other hand, lattice calculations may provide this information. As mentioned
in Section 2, the results of \cite{grossman} were interpreted exactly in the
way that they represent a non--perturbative measurement of the magnetic
screening mass $m_M \equiv (\lim_{{\bf k} \rightarrow 0} \pi_T(0,{\bf
  k}))^{1/2}$.

Given $\pi_T(k)$, in the general case one would have to
perform the Fourier--transformation (\ref{FT}), with $\tilde{E}$
replaced by
\begin{equation} \label{E}
  \tilde{E}_T \equiv \sqrt{\omega_n^2 + k_{\perp}^2 + \pi_T}~,
\end{equation}
and the subsequent integration over transverse momenta, cf.\
eq.\ (\ref{23}), numerically. However, since there is up to now no way to
obtain an analytical expression for $\pi_T$, we may as well make the
simplifying assumption that $\pi_T \equiv const.$, in order to proceed
analytically.  In this case, we anticipate (see also the discussion in the
previous section) that the only modification of the final result in comparison
to the interaction--free case will be an exponential damping of the correlator
due to the gluon mass $\sqrt{\pi_T}$.

Indeed, for constant $\pi_T$, the result (\ref{FT}) for the
Fourier--transformation still applies (with $\tilde{E}$ replaced by
$\tilde{E}_T$) and we are left with the integral over transverse momenta, which
is -- after our groundwork in the previous section -- readily expressed in
terms of ${\cal C}_2$:
\begin{equation}
  \frac{M^{(0)}}{T^5}(zT) = \frac{N_c^2-1}{16\pi}~\frac{1}{(2zT)^4}
  \left\{~{\cal C}_2(2z \sqrt{\pi_T}) + 2 \sum_{n=1}^{\infty} ~{\cal
    C}_2(2z \Omega_n^T) \right\}~, \label{m0s}
\end{equation}
where $\Omega_n^T \equiv (\omega_n^2 + \pi_T)^{1/2}$.  On
dimensional grounds $\pi_T = C T^2$, where $C$ is dimensionless. In
the absence of any other scale $C$ is independent of $T$, and thus the
right--hand side of (\ref{m0s}) is still a function of $zT$ only (as indicated
in the argument on the left--hand side).

For the following, we may assume $C \sim 1$. Then, for finite $n$ we have $4
\pi^2 n^2 T^2 \approx 40\, n^2 T^2 \gg \pi_T \sim T^2$, and consequently
$\Omega_n^T \approx 2 \pi nT$. Thus, as required from dimensional arguments, we
recover the $(zT)^{-5}$--singularity for $zT\rightarrow 0$, i.e., on scales
where neither the temperature nor the gluon mass $\sqrt{\pi_T}$ (which is
assumed to be of the order of $T$ anyway) are relevant.  In addition, the
exponential decay of the $n \neq 0$--terms for large $zT$ is similar to the
interaction--free case.  Consequently, these terms can be neglected for $zT
\geq 1/2$. The most important modification as compared to (\ref{m00}) occurs
for $n=0$. In the interaction--free case, we had $ {\cal C}_2(0) = 6$, cf.\
(\ref{u0}), leading to a {\em purely polynomial decay\/} $\sim (zT)^{-4}$ of
the correlator (\ref{m00}) for large $zT$. Now, however, due to the appearance
of the gluon mass $\sqrt{\pi_T}$, ${\cal C}_2(2 z \sqrt{\pi_T}) \sim \exp\{ -
2 z \sqrt{\pi_T} \}$ according to (\ref{uinf}), i.e., {\em an exponential
  decay\/} is superimposed onto the polynomial decay and, of course, {\em
  dominates at large $zT$}.  The corresponding slope is given by
$2\sqrt{\pi_T}$, i.e., since $m_M \equiv \sqrt{\pi_T}$, by twice the magnetic
gluon screening mass, as argued in Section 2. However, we expect that in the
region of small $zT$ the $(zT)^{-4}$--behavior and the $(zT)^{-5}$--singularity
still influence the correlator to a large extent.

For the electric correlator the combination of $k^{\mu}$ and ${\cal D}^{\mu
  \nu}$ occurring in eq.\ (\ref{e0}) reads with (\ref{D})
\begin{eqnarray} \nonumber
\lefteqn{ (k^0)^2 {\cal D}^{ij}(k) - k^0 k^j {\cal D}^{i0}(k) - k^i k^0
{\cal D}^{0j}(k) + k^i k^j {\cal D}^{00}(k) } \\ & = &
\frac{\delta^{ij} - k^i k^j / {\bf k}^2}{-k^2 + \pi_T}~k_0^2 -
\frac{-k^2}{-k^2 + \pi_L}~\frac{k^i k^j}{{\bf k}^2} \nonumber \\ & = &
\frac{-\omega_n^2}{\tilde{E}_T^2 + k_z^2}~\delta^{ij} - k^i k^j~ \left\{
\frac{1}{\tilde{E}_L^2 + k_z^2} - \frac{\omega_n^2}{k_{\perp}^2 + k_z^2}
\left( \frac{1}{\tilde{E}_T^2 + k_z^2} - \frac{1}{\tilde{E}_L^2 +
k_z^2} \right) \right\}~,  \label{expelc}
\end{eqnarray}
where we have defined $\tilde{E}_L^2 = \omega_n^2 + k_{\perp}^2
+ \pi_L$ in analogy to (\ref{E}). Again, this result is gauge invariant.

It is noteworthy that not only $\pi_L$ enters the expression for the electric
correlator, but also $\pi_T$. The reason is that there is a transverse gluon
${\cal A}^i$ involved in the definition of $E(r,0)$. One might therefore
suspect that the exponential decay of the electric correlator will be
eventually dominated by the smaller polarization function which could be
$\pi_T$, at least at high temperatures where $g\ll 1$ and $\pi_T \sim g^4
T^2 \ll \pi_L \sim g^2 T^2$.  Nevertheless, this is not the case: in eq.\
(\ref{expelc}) the respective terms are always accompanied by a factor of
$\omega_n^2$, thus they only contribute for non--zero Matsubara frequencies. As
we have seen, these terms are in general strongly suppressed for large $z$.
Furthermore, note that due to the structure of the projectors ${\cal P}^{\mu
  \nu}$ and ${\cal Q}^{\mu \nu}$, there is also a denominator $k_{\perp}^2 +
k_z^2$.

The electric correlator is given by inserting (\ref{expelc}) into (\ref{e0}).
Again, further progress is possible under the assumption that both $\pi_T$ and
$\pi_L$ are constant. As mentioned above, to one--loop order this is only true
for $k^{\mu} = (0, {\bf k})$: $\pi_T(0,{\bf k}) \equiv 0$ and
$\pi_L(0,{\bf k}) \equiv m_E^2$. The Fourier--transformation can be performed
using (\ref{FT}) and \cite{gradst}
$$
\int_{-\infty}^{\infty} \frac{{\rm d}x}{2
\pi}~e^{ixz}~\frac{1}{(a^2+x^2)(b^2+x^2)} = \frac{a~e^{-bz} -
b~e^{-az}}{2ab~( a^2- b^2 )}~,~~a,b,z >0~.
$$
We rearrange the final result in terms proportional to the various
exponential decay factors,
\begin{eqnarray}
  \lefteqn{ E^{(0)} (z) = \frac{N_c^2 - 1}{16 \pi}~T
    \sum_{n=-\infty}^{\infty} \int_0^{\infty}{\rm d}k_{\perp}
    k_{\perp}~\left\{~\omega_n^4~\left(~\left[ \frac{1}{\tilde{E}_T^2} +
    \frac{\tilde{E}_T^2}{(\Omega_n^T)^4} \right]~e^{-2 \tilde{E}_T z}
  \right. \right.} \nonumber \\ & & \left.+~~\frac{\pi_T - \pi_L}{ (\Omega_n^T
  \Omega_n^L)^2}~\frac{2 k_{\perp}}{\tilde{E}_T}~\left[ 1 +
\frac{k_{\perp}^2}{(\Omega_n^T)^2} \right]~e^{-(k_{\perp} + \tilde{E}_T)
  z} + \frac{(\pi_T - \pi_L)^2}{(\Omega_n^T
  \Omega_n^L)^4}~k_{\perp}^2~e^{-2 k_{\perp} z} \right) \label{e0s} \\ &\!
+ & \left. \!\!\! \omega_n^2~\frac{\pi_L}{(\Omega_n^T
\Omega_n^L)^2}~\frac{2 k_{\perp}^2}{\tilde{E}_L}
\left( \frac{\pi_T-\pi_L}{(\Omega_n^L)^2}~
k_{\perp}~e^{-(k_{\perp}+\tilde{E}_L) z} +
 \tilde{E}_T~e^{-(\tilde{E}_T +
  \tilde{E}_L)z} \right) +
\frac{\pi_L^2}{(\Omega_n^L)^4}~\frac{k_{\perp}^4}{\tilde{E}_L^2}~e^{
  -2\tilde{E}_L z} \right\}~.\nonumber
\end{eqnarray}
Here we have also introduced $\Omega_n^L \equiv (\omega_n^2 +
\pi_L)^{1/2}$. The integration over transverse momenta can be done
analytically, but the result is extremely unwieldy. In fact, most of the terms
are rapidly suppressed for increasing $z$, since they are accompanied by terms
$\sim \omega_n^2~\exp\{ - z\Omega_n^{T,L} \}$. As we will see, for a
qualitative discussion of the correlator it is sufficient to retain only the
last terms in the second and the third line of eq.\ (\ref{e0s}),
\begin{eqnarray}
  \int_0^{\infty} {\rm d}k_{\perp}
  k_{\perp}~k_{\perp}^2~e^{-2k_{\perp}z} & = & \frac{1}{(2z)^4}~{\cal
    C}_2(0)~, \label{poly} \\ \int_0^{\infty} {\rm d}k_{\perp}
  k_{\perp}~\frac{k_{\perp}^4}{\tilde{E}_L^2}~e^{-2\tilde{E}_L z} & = &
  \frac{1}{(2z)^4}~{\cal C}_2(2z\Omega_n^L)~. \label{integ}
\end{eqnarray}
For small $z$, the Matsubara sum over the terms (\ref{integ})
provides the $z^{-5}$--singularity expected from dimensional arguments.  For
$zT \geq1$ and finite $n$, the term (\ref{integ}) becomes irrelevant, because
it is exponentially suppressed $\sim \exp \{ - 2 z \Omega_n^L \} \approx \exp
\{ - 4 \pi n zT \}$ on account of (\ref{uinf}).  For $n=0$, however, (a) it is
the only term in (\ref{e0s}) which does not vanish and (b) it exhibits the
weakest {\em exponential\/} decay $\sim \exp \{ - 2z\sqrt{\pi_L} \}$ of all
terms in (\ref{e0s}).  The slope of this decay is twice the electric screening
mass $m_E\equiv \sqrt{\pi_L}$. The term (\ref{poly}) is only {\em
  polynomially\/} decaying ($\sim z^{-4}$) since ${\cal C}_2(0)=6$, and
will thus eventually dominate the long--distance behavior of the correlator
(\ref{e0s}), provided $\pi_T \neq \pi_L$ (which appears as a prefactor to
(\ref{poly}) in (\ref{e0s})). Note that this term arises from the
Fourier--transformation of the denominator $k_{\perp}^2 + k_z^2$ in
(\ref{expelc}) which in turn originates from the specific structure of the
projectors in eq.\ (\ref{D}). For $zT \geq 1$ the approximate form of the
electric correlator is
\begin{equation} \label{e0sb}
  \frac{E^{(0)}}{T^5}(zT) \approx \frac{N_c^2-1}{16
    \pi}~\frac{1}{(2zT)^4} \left\{~{\cal C}_2(2z\sqrt{\pi_L}) +
  12~(\pi_T - \pi_L)^2 \sum_{n=1}^{\infty}~\frac{\omega_n^4}{(\Omega_n^T
    \Omega_n^L)^4} \right\}~.
\end{equation}
In Fig.\ 3a we show the electric and magnetic correlator,
normalized to $T^5$, as a function of $zT$ for $\pi_L =(1.4\, T)^2,\,
\pi_T=(2.9\, T)^2$ in comparison to the free case. This choice for the
polarization functions is motivated by the results of Ref.\ \cite{grossman},
where it was found that the electric correlator decays with the rate $2\, m_E
\equiv 2 \sqrt{\pi_L} \approx 2.8\, T$ and the magnetic correlator with the
rate $2\, m_M \equiv 2 \sqrt{\pi_T} \approx 5.8\, T$.

One observes that the power--law behavior found in the interaction--free case
dominates both correlators up to $zT \approx 2$. For larger $zT$ the
exponential decay prevails for the magnetic while for the electric correlator
an anomalous power--law decay is observed due to the second term in
(\ref{e0sb}).  This is shown in detail in Fig.\ 3b, where the various
contributions to the electric correlator are plotted separately.  The precise
value of $zT$ at which the power--law term dominates over the exponentially
decaying first term in (\ref{e0sb}) depends on $\pi_T$ and $\pi_L$.  This
anomaly is an artefact from our simplifying assumption that the polarization
functions do not vanish for large ${\bf k}$ and non--zero $\omega_n$. In
general we expect that for large $zT$ the electric correlator decays
exponentially on the scale $2\, m_E$.

\section{Lattice correlators  in the weak--coupling limit}

Due to the periodicity of the lattice, a simulation can follow the decay of any
correlator only up to $z/L =1/2$, where $L=aN_{\sigma}$ is the spatial
extension of the lattice.  This point corresponds, for any lattice with
$N_{\sigma} = 4\, N_{\tau}$, to $zT =2$. What is surprising is that in the
lattice simulations of \cite{grossman} there is no hint of the power--law
behavior which we have found to dominate the correlators for $zT \leq 2$ in our
above continuum analysis.  As we will show in the next section, this could be
due to the use of fuzzed links \cite{private}.

In this section we quantify effects arising from the discretization and the
finite lattice size without distortion from additional signal enhancement
prescriptions. For this purpose it is sufficient to only consider the magnetic
correlator.  The transcription of eq.\ (\ref{m0}) onto a finite lattice
proceeds along the lines given in \cite{heller}. After switching to the
Euclidean metric, the main difference between the continuum and the lattice
formulation is (a) that the gluon fields live in the middle of a link
\begin{equation} \label{49}
  A_{\mu}^b(x) = \frac{1}{N_{\sigma}^3 N_{\tau} a^4}~\sum_k
  e^{ik_{\alpha} (x + a{\bf e}_{\mu}/2)_{\alpha}}~A_{\mu}^b(k)~,
\end{equation}
where ${\bf e}_{\mu}$ is the lattice unit vector pointing
in $\mu$--direction, and (b) that partial derivatives are replaced by
forward finite differences,
\begin{equation}
\partial_{\nu} A_{\mu}^b(x)~\rightarrow~\frac{ A_{\mu}^b(x + a {\bf
    e}_{\nu}) - A_{\mu}^b(x)}{a}~.
\end{equation}
The momenta in (\ref{49}) fulfill
\begin{eqnarray*}
  k_i & = & \frac{2 \pi
    j_i}{aN_{\sigma}}~,~~j_i=0,...,N_{\sigma}-1~,~~i=x,y,z~,\\ k_0 & = &
  \frac{2 \pi j_0}{a N_{\tau}}~,~~j_0 = 0,...,N_{\tau}-1~.
\end{eqnarray*}
The free lattice action closely resembles the continuum action (\ref{action}),
\begin{equation}
  S_0[A] = - \frac{1}{2 N_{\sigma}^3 N_{\tau} a^4} \sum_k
  A_{\mu}^b(-k)~{\cal D}_{0, \mu \nu}^{-1}(k)~A_{\nu}^b(k)~,
\end{equation}
with
\begin{equation} \label{lattprop}
  {\cal D}^{-1}_{0,\mu \nu} (k) = \delta_{\mu \nu}~\left( m^2 +
  \frac{4}{a^2} \sum_{\rho=\tau}^z \sin^2 \left[ \frac{k_{\rho}a}{2}
\right] \right)~,
\end{equation}
where we have introduced a mass term. On the one hand, this
allows to regularize any singularities due to constant field configurations.
Usually, these singularities are avoided by omitting the $k_0
=k_x=k_y=k_z$--term in momentum sums \cite{heller}. We will see that this is
not necessary here: even for $m=0$ such a term is finite in the correlator
(see below). On the other hand, allowing for a finite $m$ from the outset
provides a simple way to include a screening mass in the final expression.  In
the following, we will also make use of the relation
\begin{equation}
  \left\langle A_{\mu}^b(k)~A_{\nu}^c(l) \right\rangle_0 =
  \delta_{bc}~\delta^{(4)}_{k,-l}~N_{\sigma}^3 N_{\tau} a^4~{\cal D}_{0,
    \mu \nu}(k)~.
\end{equation}
Given these facts, it is straightforward to derive the
magnetic correlator
\begin{eqnarray} \label{lattmgen}
  \lefteqn{ M^{(0)}_0(r,0) \equiv \left\langle {\rm Tr}~{\cal
      F}_{xy}^2(r)~{\rm Tr}~{\cal F}_{xy}^2(0) \right\rangle_{0,c} =
    \frac{N_c^2-1}{2} \left(~\frac{1}{N_{\sigma}^3 N_{\tau} a^4}~\frac{4}{a^2}
    ~\sum_k~ e^{ik \cdot r} \right. } \\ & \times & \left. \left\{
  \sin^2 \left[ \frac{k_x a}{2}\right]~{\cal D}_{0, yy}(k) - 2 \sin
  \left[ \frac{k_x a}{2}\right] \sin \left[ \frac{k_y a}{2}\right]~{\cal
    D}_{0, xy}(k) + \sin^2 \left[\frac{k_y a}{2}\right]~{\cal
    D}_{0,xx}(k) \right\} \right)^2~.\nonumber
\end{eqnarray}
The main difference as compared to (\ref{m0}) is that $k^i$ is
replaced by $2 \sin[k_ia/2]$, as usual for bosons on the lattice.  Summing over
the $\tau-,x-$, and $y-$sites at constant $z$ in order to project onto zero
momentum and employing (\ref{lattprop}) we obtain
\begin{equation} \label{lattm}
  M^{(0)}_0(z) (aN_{\tau})^5 =
  \frac{N_c^2-1}{2}~\frac{N_{\tau}^4}{N_{\sigma}^2} \sum_{k_0,k_x,k_y}
  \left( \frac{4}{N_{\sigma}}~\sum_{k_z}~e^{ik_z z}~\frac{ \sin^2 [k_x
    a/2] + \sin^2 [k_y a/2] }{(ma)^2 + 4 \sum_{\rho} \sin^2[k_{\rho}
    a/2] } \right)^2~.
\end{equation}
The Fourier--transformation can be done using the
identity
\begin{equation}
  \frac{1}{N_{\sigma}} \sum_{j_z=0}^{N_{\sigma}-1} \frac{\cos [2 \pi j_z
    z/aN_{\sigma}]}{ \omega^2 + \sin^2 [\pi j_z/a N_{\sigma}] } \equiv
  2~\frac{ \cosh \left[ 2 \hat{\omega} (N_{\sigma}/2 - z/a) \right] }{
    \sinh [ 2 \hat{\omega}]~\sinh [N_{\sigma} \hat{\omega}] }~,
\end{equation}
which can be proven by contour integration.  Here
$\hat{\omega} = \ln [ \omega + \sqrt{\omega^2+1} ]$ and
$$
\omega^2 \equiv \left( \frac{ma}{2} \right)^2 + \sum_{\rho = \tau,x,y} \sin^2
\left[ \frac{k_{\rho}a}{2} \right]~.
$$
The reason why there is no singularity in (\ref{lattm}) in the case $m=0$
due to constant field configurations ($k_0 = k_x = k_y = k_z =0$) is that the
term $\sin^2 [k_x a/2] + \sin^2 [k_y a/2]$ regulates this singularity.
Nevertheless, the limit $m \rightarrow 0$ is subtle: putting $m=0$ {\em
  from the outset}, the zero momentum term is {\em finite}, while performing
the limit $m \rightarrow 0$ {\em after\/} calculating the correlator, this
term {\em vanishes}. Since the theory is not well-defined in the first case, we
will always use the second prescription when considering the case $m=0$ in
the following.

In the case of a dressed gluon propagator one would in principle have to use
the lattice analogues for the decomposition (\ref{D}) and for the polarization
functions $\pi_{T,L}$ \cite{elze}.  However, for an estimate of lattice effects
and the general behavior of the correlator on the lattice it is certainly
sufficient to simply assume a finite constant mass $m \equiv \sqrt{\pi_T}$ in
the above expression (\ref{lattm}). This assumption is supported by a
comparison of (\ref{lattmgen}) and the continuum analogue (\ref{m0}) with
(\ref{35}).

The resulting correlator is shown in Fig.\ 4a for $\pi_T = 0$ and $\pi_T =
(2.9\, T)^2$ in comparison to the corresponding continuum expression. One
observes that the lattice ultraviolett cut--off $a^{-1}$ regulates the
$z^{-5}$--singularity at the origin, but that otherwise, on the $4 \times
16^3$--lattice used in \cite{grossman}, the continuum limit is quite well
reproduced.  This, however, implies that the power--law behavior still
influences the decay of the correlator.  This has important consequences for
the extraction of a screening mass.  A fit with a simple exponential $\exp\{ -
\mu(z) z \}$, or, on a finite lattice with a hyperbolic cosine $\cosh [ \mu(z)
(aN_{\sigma}/2 - z)]$ \cite{grossman} (due to the lattice symmetry centered at
$z/L = 1/2$), will necessarily lead to local masses $\mu(z)$ which are {\em
  larger\/} than $2\sqrt{\pi_{T}}$ (or $2\sqrt{\pi_L}$ in the electric case).
Thus, screening masses extracted from these local masses via the simple
relation $\mu(z) = 2\, m_{E,M}$ {\em overestimate\/} the actual values.

To quantify this we extract local masses from the correlators via
\begin{equation} \label{fit}
  \frac{M_0^{(0)}(z)}{M_0^{(0)}(z-a)} = \frac{\cosh [ \mu(z) (aN_{\sigma}/2 -
    z) ]}{\cosh [ \mu(z) (aN_{\sigma}/2 - (z-a)) ]}~,
\end{equation}
which is the same procedure as used in Ref.\ \cite{grossman}. If
this prescription worked correctly, the local mass should be two times the
gluon mass, i.e., $\mu(z)a \equiv 2\, ma = 2 \sqrt{\pi_T/T^2}/N_{\tau} = 1.45 =
\mu_{real}a$ ($N_{\tau}=4$ and $\pi_T=(2.9\, T)^2$). However, as one observes
in the first row of Table 1 (see also Fig.\ 5a), even at $z/a=6$ the local
mass overestimates the actual value by 35\%~!

However, our results are not yet directly comparable to that of
\cite{grossman}.  First of all, in that work sources and sinks of the
correlators are plaquettes with a minimum edge length $a$, while (\ref{lattm})
is essentially a point--to--point correlator. In addition, {\em fuzzed links\/}
were used to enhance the signal at large $z$ \cite{fuzz}. This technique
effectively enlargens sources and sinks to the size of several plaquettes and
also suppresses contributions of higher excited states at small $z$
\cite{potential}, thus having an effect similar to wall sources commonly
employed in the calculation of hadronic correlation functions. In the next
section we investigate a model for fuzzing and apply it to our perturbative
correlators.

To conclude this section we remark that excited states are in our picture
essentially states of {\em non--zero\/} transverse momentum and Matsubara
frequency in the gluon loop in Figs.\ 1a,d. The summation over these excited
states is nothing but the integration over transverse momenta and summation
over Matsubara frequencies in (\ref{23}) or (\ref{e0s}).  In this sense, the
common notion that higher excited states enhance the point--to--point
correlator at small distances is just reflected by our observation that the
integration over transverse momenta and the summation over Matsubara
frequencies leads to polynomial enhancement for small $z$.
 \newpage
\begin{center}
{\small \bf Table 1} \\ ~~ \\
\begin{tabular}{|c||c|c|c|c|c|c||c|c|c|} \hline
$z/a$    &  1 & 2 & 3 & 4 & 5 & 6 & $\mu_{real}a$ & $L_{\tau}$ &
                                        $L_{\sigma}$  \\ \hline \hline
         & 4.09 & 3.58 & 2.95 & 2.45 & 2.15 & 1.95 & 1.45 & 0 & 0
                                                        \\ \cline{2-10}
         & 3.28 & 2.86 & 2.50 & 2.24 & 2.06 & 1.91 & 1.45 & 1 & 1
                                                        \\ \cline{2-10}
 $\mu(z)a$ & 2.91 & 2.50 & 2.25 & 2.08 & 1.97 & 1.85 & 1.45 & 2 & 2
                                                        \\ \cline{2-10}
         & 2.80 & 2.29 & 2.01 & 1.89 & 1.81 & 1.72 & 1.45 & 2 & 4
                                                        \\ \cline{2-10}
         & 2.79 & 2.25 & 1.92 & 1.77 & 1.70 & 1.62 & 1.45 & 2 & 8
                                                      \\  \hline \hline
         & 3.83 & 2.93 & 1.87 & 1.26 & 0.96 & 0.77 & 0    & 0 & 0
                                                        \\ \cline{2-10}
         & 2.83 & 2.11 & 1.52 & 1.16 & 0.93 & 0.76 & 0    & 1 & 1
                                                        \\ \cline{2-10}
 $\mu(z)a$ & 2.30 & 1.69 & 1.31 & 1.07 & 0.89 & 0.74 & 0    & 2 & 2
                                                        \\ \cline{2-10}
         & 1.97 & 1.31 & 1.05 & 0.92 & 0.81 & 0.71 & 0    & 2 & 4
                                                        \\ \cline{2-10}
         & 1.80 & 1.10 & 0.89 & 0.81 & 0.74 & 0.66 & 0    & 2 & 8
                                                             \\  \hline
\end{tabular}
\end{center}

\section{Extended correlators}

Fuzzed links are obtained by replacing each link by a sum of products of two or
more links in the vicinity of the old link \cite{fuzz}. This roughly doubles
the size of plaquettes at each new level of fuzzing and thus smears out the
sources and sinks of the plaquette correlators.

In our continuum analysis, we model fuzzing by introducing form factors,
$\Phi(\Delta r)$ and $\Phi'(\Delta r')$, for the smeared sink and source,
respectively, and study the extended correlator
\begin{eqnarray} \label{smear}
  {\cal M}(z) & \equiv & \frac{T}{L^2} \int {\rm d}^3r_1~{\rm d}^3r_2~{\rm
    d}^3r_1'~{\rm d}^3r_2' ~\Phi(\Delta r)~\Phi'(\Delta r')~{\cal
    M}(r_1,r_2,r_1',r_2')~, \\ {\cal M}(r_1,r_2,r_1',r_2') & \equiv &
  \left\langle {\rm Tr}[{\cal B}_z(r_1) {\cal B}_z(r_2)]~{\rm Tr}[{\cal
    B}_z(r_1'){\cal B}_z(r_2')] \right\rangle_c~,\label{smear2}
\end{eqnarray}
where $\Delta r \equiv r_1 - r_2,\, \Delta r' \equiv r_1' -
r_2'$ ($\Delta z = \Delta z' =0$) and ${\rm d}^3r_i \equiv {\rm d}\tau_i\, {\rm
  d}x_i\, {\rm d}y_i,\, i=1,2$, and similarly for ${\rm d}^3r_i'$.  Note that,
analogous to fuzzing \cite{grossman,fuzz}, smearing is limited to the
transverse and temporal directions with $z'_1=z'_2=0$ and $z_1=z_2=z$ held
fixed.  Equation (\ref{smear}) reduces to (\ref{project}) for the choice
$\Phi(\Delta r) = \delta^{(3)}(\Delta r)$, $\Phi'(\Delta r') =
\delta^{(3)}(\Delta r')$.  The prescription (\ref{smear}) is chosen such that
(a) the only modification of the one--loop contribution (\ref{23}) is the
additonal appearance of the Fourier--transformed form factors
$\tilde{\Phi}(-\omega_n,-{\bf k}_{\perp})$ and $\tilde{\Phi}'(\omega_n, {\bf
  k}_{\perp})$ and (b) its gauge invariance is preserved due to (\ref{free}).

It is obvious that with suitable form factors one can suppress the contribution
of large $k_{\perp}$ and $n$. Thus, our model correctly simulates the effect of
fuzzing in that the small--$z$ enhancement of the correlation function due to
higher excited states can be reduced, which leaves the exponential decay less
distorted.  Diagramatically, the above prescription replaces Fig.\ 1d with
point source and sink with Fig.\ 1f, where the two dressed gluons are produced
and absorbed from an extended source and sink, respectively.

For example, {\em wall sources\/} correspond to the special choice $\Phi'
(\Delta r') = T/L^2 \equiv const.$, $\Phi (\Delta r) = \delta^{(3)}(\Delta r)$.
This corresponds to $\tilde{\Phi}'(\omega_n,{\bf k}_{\perp}) =
\delta_{n,0}~\delta^{(2)}_{{\bf k}_{\perp},0}\, ,\, \tilde{\Phi} = 1$, which
has the maximum effect of suppressing {\em all\/} higher transverse momenta and
Matsubara frequencies. Unfortunately, wall sources are not applicable in our
case, since the contribution of zero transverse momentum has zero measure in
the integral in (\ref{23}) and would thus lead to a trivial result for the
correlator.

We now investigate the influence of smeared sources and sinks on the lattice
correlator (\ref{lattm}). To this end, eq.\ (\ref{smear}) is transcribed onto
the lattice using standard discretization prescriptions. In \cite{grossman} the
sources and sinks prior to fuzzing are plaquettes of edge length $a$. Fuzzing
links up to fuzzing level $l$ increases the {\em average\/} edge length of the
plaquette source and sink to $2^l a$ \cite{fuzz}.  Consequently, the sources
and sinks have on the average the volume $L_{\tau}a \times (L_\sigma a)^2$
where $L_{\tau} \equiv 2^{l_\tau},\, L_{\sigma} \equiv 2^{l_\sigma}$.  Thus, we
choose as form factor
\begin{equation} \label{phi}
  \Phi(r) \equiv  \frac{1}{a^3(L_{\tau}+1)(L_{\sigma}+1)^2}~
  \sum_{n_{\tau}=0}^{L_{\tau}}~
  \sum_{n_x,n_y=0}^{L_{\sigma}}~\delta_{\tau,n_{\tau}a}~ \delta_{x,n_x
    a}~\delta_{y,n_y a}~,\label{form}
\end{equation}
and the same for $\Phi'$.  This corresponds to a homogeneous
smearing of the source and sink over a cube of volume $L_{\tau}a \times
(L_{\sigma}a)^2$. Clearly many other possibilities for form factors are
possible, but (\ref{form}) is the most simple choice and will be adequate to
prove our point, namely, that the extracted screening
masses are extremely sensitive to the chosen smearing prescription and
that smearing improves the extraction of screening masses only in the
case when those are large compared to $T$. For small
screening masses smearing will be seen to introduce uncontrolled errors.

The fuzzing level used in \cite{grossman} is $l_{\tau,\sigma}=\log_2
N_{\tau,\sigma} -1$, i.e., $l_{\sigma}=3$ and $l_{\tau}=1$ for the $4 \times
16^3$--lattice under consideration. Consequently, $L_{\sigma} =8$ and
$L_{\tau}=2$. However, in order to study systematically the influence of our
``fuzzing'' prescription, we vary $L_{\sigma}$ from 0 to 8 and $L_{\tau}$ from
0 to 2 (see Table 1 and Figs.\ 4b,c, and 5a).  Note that since sources and
sinks of the correlators studied in \cite{grossman} are elementary plaquettes
of edge length $a$ prior to fuzzing, for instance an ``unfuzzed''
$P_s$--correlator should actually not be compared to our magnetic
point--to--point correlator of Section 6, but rather to an extended correlator
with $L_{\tau}=0, \, L_{\sigma}=1$.  The Fourier--transform of (\ref{phi}) is
\begin{eqnarray}
  \tilde{\Phi}(\omega_n,{\bf k}_{\perp})& = & a^3~\sum_{\tau,x,y}~
  e^{-i(\omega_n \tau + k_x x + k_y y)}~\Phi(r) \nonumber \\
  & = &  \frac{1}{(L_{\tau}+1)(L_{\sigma}+1)^2}~\sum_{n_{\tau}=0}^{L_{\tau}}~
  \sum_{n_x,n_y=0}^{L_{\sigma}}~e^{-ia(\omega_n n_{\tau} + k_x n_x + k_y
    n_y)}~.\label{tilphi}
\end{eqnarray}
The result (\ref{lattm}) is modified in that the modulus squared
of (\ref{tilphi}) appears additionally under the sum over $k_0,k_x,k_y$.  Using
\cite{gradst5}, one obtains the simple analytical expression
\begin{equation}
  |\tilde{\Phi}(\omega_n,{\bf k}_{\perp})|^2 = \left( \frac{ \sin [
    a\omega_n(L_{\tau}+1)/2]}{(L_{\tau}+1) \sin[a \omega_n/2]}~\frac{
    \sin [ ak_x (L_{\sigma}+1)/2]}{(L_{\sigma}+1) \sin[ak_x/2]}~\frac{
    \sin [ ak_y (L_{\sigma}+1)/2]}{(L_{\sigma}+1) \sin[ak_y/2]}
\right)^2~.
\end{equation}
The correlator is displayed in Fig.\ 4b for $\pi_T=(2.9\, T)^2$
and in Fig.\ 4c for $\pi_T=0$ for different degrees of smearing.  One clearly
observes a suppression of the polynomial enhancement at small $z$, which is
proportional to the degree of smearing of sources and sinks.  This reflects the
sensitivity of the correlators and hence the extracted screening masses to the
choice of the form factor and thus to the particular fuzzing prescription. In
the case of maximal smearing $L_{\sigma}=8,\, L_{\tau}=2$ the decay of the
correlator for $\pi_T = (2.9\, T)^2$ is very close to exponential, which is
confirmed by the fact that the local masses are nearly constant for $2 \leq z/a
\leq 6$.  This can be seen from Table 1 and Fig.\ 5a.  However, while the
reduction of the error due to smearing is impressive for small $z$, the
overestimate is still approximately $12\%$ at $z/a=6$.

Reversing this argument, one could ask for the value of the {\em real\/}
screening mass that would lead to an observed {\em local\/} mass near the value
$\mu(z)a \approx 1.45$. To this end, consider first the local masses extracted
from the {\em free\/} correlator (Fig.\ 4c). One observes (cf.\ Table 1 and
Fig.\ 5a) that the sensitivity of the local masses to the degree of smearing
(and thus the fuzzing prescription) is even more pronounced than in the finite
mass case, especially for small $z$. However, the striking feature is that,
although the {\em theoretically assumed screening mass is zero}, the {\em local
  mass\/} extracted via the exponential--type fit (\ref{fit}) is {\em finite}~!
This is independent of the degree of smearing, although more strongly smeared
sources and sinks give smaller values for the local mass. Nevertheless, the
local mass extracted via (\ref{fit}) fails completely to approximate the real
mass if the latter is small compared to $T$.

To compare with the results of \cite{grossman} we have to bear in mind that our
perturbative analysis relies on the assumptions that (a) plaquette correlators
can be approximated by energy density correlators, (b) the main contribution to
the energy density correlators is the exchange diagram of Fig.\ 1d, and (c)
that our smearing prescription accurately models fuzzing as employed in
\cite{grossman}.  The limitations of (a) and (b) have already been discussed in
Section 2 and 3.  Assumption (c) could be questionable in view of the following
fact: the exact shape of the source and the sink of the correlator, i.e., the
fuzzing or smearing prescription, is irrelevant only as long as there is
sufficiently large overlap with the wavefunction of the state the mass of which
is to be measured.  This holds in general only for bound states. In our case,
however, the correlator is supposed to be given by the exchange of two
independently propagating, dressed gluons. We have already shown (cf.\ Figs.\
4b,c, and 5a) that in this case the correlator is very sensitive to the chosen
smearing (fuzzing) prescription.

However, assuming the validity of our assumptions, it is surprising that on the
average the local masses $\mu(z)$ in the case $\pi_T=0$ are close to the value
obtained in \cite{grossman} for the magnetic correlator~!  In order to get the
value of $\mu(z)a \approx 1.45$ at $z/a=3$ (which was considered to give the
best estimate for the mass determining the decay of the magnetic correlator in
\cite{grossman}), we have to assume $\pi_T$ to be at most of the order of $(2\,
T)^2$ (cf.\ Fig.\ 5b). On the average, $\pi_T \approx T^2$ (corresponding to
$\mu_{real}a=0.5$) can be best reconciled with the data of \cite{grossman}.
This, in turn, means that the local masses may easily overestimate the real
masses by at least a factor of two.

However, a more detailed comparison of our results with the lattice data
reveals further problems with this interpretation.  In our calculation, the
general behavior of the local masses still reflects the remnant enhancement at
small $z$, giving rise to larger $\mu(z)$ at small $z$. This trend is not seen
in the lattice data and could indicate that our form factor ansatz does not
accurately model fuzzing as employed in \cite{grossman}.  We suggest that a
lattice test of the dependence of the local masses on the level of fuzzing
could help in clarifying the situation.

On the other hand, if our fuzzing model works accurately, we are lead to the
conclusion that either the plaquette correlators may not be approximated by
energy density correlators or that the latter are not dominated by the
two--gluon exchange graph of Fig.\ 1d.  If the first possibility is true, then
finite size effects are still too large to allow for a simple interpretation of
the results of \cite{grossman} in terms of continuum physics. In the second
case, a continuum analysis which improves and corrects the simple picture
investigated in the present work will be extremely difficult, since one does
not a priori know which multi--gluon exchange diagrams yield the dominant
contribution to the energy density correlators. However, in view of recent
measurements of the spatial string tension \cite{string} one might speculate
that the magnetic correlator is dominated by a bound state, as represented by
Fig.\ 1e.  To confirm this, however, would require lattice data with much
better statistics which is extremely difficult to obtain for the plaquette
operators studied in \cite{grossman}. In any case, if this picture is true, the
interpretation of the screening mass extracted from the decay of the magnetic
correlator as twice the gluon magnetic screening mass is not possible.
Moreover, this indicates the breakdown of the simple perturbative picture of
the gluon plasma, in accordance with the conclusions of \cite{dhr}.

Finally we comment on the electric correlator. Given the fact that in the
continuum it is not very different from the magnetic, we may use (\ref{lattm})
as a rough approximation also for the electric correlator. However, in that
case the smaller local electric masses found in \cite{grossman} ($ma \approx
0.8$, cf.\ Fig.\ 5b), can only be accounted for in our model by assuming that
the actual electric screening mass is essentially zero!  This would be in clear
contradiction to the screening mass extracted from Polyakov--loop correlations
\cite{grossman} (unless they also suffer from ambiguities).

We conclude that independent measurements of especially the non--perturbative
magnetic screening mass are needed to better pin down this physically
interesting and phenomenolo\-gi\-cal\-ly relevant quantity \cite{private2}.
\\ ~~ \\
\noindent
{\bf Acknowledgments}
\\ ~~ \\
We thank R.D.\ Pisarski for bringing
this problem to our attention and T.\ Biro, G.\ Boyd, N.\
Christ, F.\ Karsch, V.\ Koch, R.\ Mawhinney, A.\
Sch\"afer, and M.\ Thoma for stimulating discussions. We are also indebted
to U.\ Heller, F.\ Karsch, and R.\ Mawhinney for a critical reading of the
final version of the manuscript and helpful comments. \\ ~~ \\

\newpage
\noindent
{\bf Figure Captions:}
\\ ~~ \\
{\bf Fig.\ 1:}
Topologically distinct
graphs contributing to the energy density correlator: (a) one--loop, two--gluon
exchange diagram with free gluon propagators, (b,c) two-- and three--loop
diagrams with free gluon propagators, (d) one--loop, two--gluon exchange
diagram with dressed gluon propagators, (e) generic ladder diagram representing
propagation of a bound gluon pair, and (f) one--loop, two--gluon exchange with
smeared source and sink.
\\ ~~ \\
{\bf Fig.\ 2:}
(a) The magnetic correlator
in the interaction--free case normalized to $T^5$ as a function of $zT$ (full
line) and the contribution of the zero Matsubara mode (dotted line). (b) The
normalized magnetic (dotted curve) and electric correlator (full line) as
functions of $zT$.
\\ ~~ \\
{\bf Fig.\ 3:}
(a) The magnetic (short--dashed
curve) and electric correlator (long--dashed curve) for $\pi_L = (1.4\, T)^2$
and $\pi_T = (2.9\, T)^2$ in comparison to the interaction--free result of
Fig.\ 2b. (b) Different contributions to the electric correlator: the dotted
curve corresponds to the contribution of the zero mode, while the full line
represents the full result. The exponentially damped terms with $n \neq 0$
constitute the difference between these curves at small $zT$, the anomalous
polynomially decaying term the difference for large $zT$.
\\ ~~ \\
{\bf Fig.\ 4:}
(a) The magnetic correlator for the interaction--free case in the
continuum (long--dashed curve: $\pi_T = 0$, dotted curve: $\pi_T =
(2.9\, T)^2$) in comparison to a calculation on a $4 \times 16^3$--lattice
(squares: $m=0$, circles: $m=2.9\, T$). (b) The magnetic correlator on the
lattice for $m=2.9\, T$ and different degrees of smearing. The solid curve
corresponds to point sources and sinks ($L_{\sigma}=L_{\tau}=0$), the dashed
curve to smearing with $L_{\sigma} =L_{\tau}=1$, the dashed--dotted curve to
$L_{\sigma}=L_{\tau}=2$, the long--dashed curve to $L_{\sigma}=4,\,
L_{\tau}=2$, and the dotted curve to $L_{\sigma}=8,\, L_{\tau}=2$. (c) The
same as in (b) for $m=0$.
\\ ~~ \\
{\bf Fig.\ 5:}
(a) Local masses extracted
via (\ref{fit}) from the magnetic correlators of Figs.\ 4b,c. The upper set of
curves is for $m=2.9\, T$, the lower for $m=0$. (b) Local masses extracted
from the magnetic correlator with maximal smearing ($L_{\sigma}=8,\,
L_{\tau}=2$). The curves are labelled with the real gluon mass entering the
calculation of the correlator.  Black squares are local masses extracted from
data \cite{grossman} for the magnetic correlator, open circles are local masses
corresponding to the electric correlator.

\begin{thebibliography}{99}
\bibitem{footn6} The bare mass is even smaller and can be
completely neglected for this consideration.
\end{thebibliography}
\end{document}